# Personality Dimensions and Temperaments of Engineering Professors and Students – A Survey

Arif Raza, Zaka-ul-Mustafa and Luiz Fernando Capretz

**Abstract----**This research work aims to study personality profiles and temperaments of Pakistani software engineering professors and students. In this survey we have collected personality profiles of 18 professors and 92 software engineering students. According to the Myers-Briggs Type Indicator (MBTI) instrument, the most prominent personality type among professors as well as among students is a combination of Introversion, Sensing, Thinking, and Judging (ISTJ). The study shows ITs (Introverts and Thinking) and IJs (Introverts and Judging) are the leading temperaments among the professors. About the students' data, the results of the study indicate SJs (Sensing and Judging) and ISs (Introverts and Sensing) as the dominant temperaments.

**Keywords----** Human Factors in Software Design, Process metrics, Software Engineering Process, Statistical methods,

──────── ◆ ─────────

## 1 INTRODUCTION

Different perspectives of human factors in software engineering have been explored using MBTI. These perspectives include human factors in different phases of the software life cycle, the effect of team work in software development, or the correspondence between personality profiles and tasks. Professors and students are the corner stones of any education system. Software engineering education, its application and success rely on these communities as well. The acquaintance among personalities of these communities is a way to lead smooth and valuable education.

Many studies have been carried out to exhibit personality profiles of software professionals using MBTI tool [1]. However, significant data is not available related to South-Asian software professionals. To acquire and study personality profiles of faculty members and students is imperative to develop software engineering education in this region.

The goal of this paper is to identify and compare personality types and temperaments of software engineering professors and students. Consequently, in this study, 110 Pakistani software engineers, including students and professors of the National University of Sciences and Technology, Islamabad, Pakistan are surveyed.

## 2 THE MYERS-BRIGGS TYPE INDICATOR (MBTI)

Number of psychological instruments are available and in vogue for career counseling and behavior prediction.

MBTI is one of the most popular tools used in organizations for the classification of personality types [2]. It has also been used to understand individual learning styles and preferences in inspiration. The MBTI has four dimensions of preferences, which describe a specific personality. Each dimension has two opposite pairs: Extroversion - Introversion, Sensing - Intuition, Feeling - Thinking, and Perceiving - Judging. As a result, sixteen typical personality types are defined by using combination of these four distinct types. An individual can possibly use all eight preferences in each of the four pairs; however in general, every person has one dominant dimension in his/her personality. The scales are briefly described below:

### 2.1 Extroversion (E) – Introversion (I)

Extroverts prefer to communicate with other people by focusing on outer world of people and things, whereas introverts choose to work independently by focusing on inner world of ideas and emotions.

### 2.2 Sensing (S) - Intuition (N)

This dimension is about the way people gain information. Sensing people trust on their experience and tend to focus on facts they can count on, while intuitive individuals are more focused on their creativity, insight and new potential of events.

### 2.3 Thinking (T) - Feeling (F)

The third dimension is about the way people take decisions in life. Thinking individuals are cool headed, prefer clearly defined tasks and have a logical and analytical reasoning to make decisions, whereas feeling people are warm hearted, consider harmonious working relationship important and have a sensitive approach.

### 2.4 Judging (J) - Perceiving (P)

Judging type likes to follow a schedule, prefers to have things settled and does not like too much spontaneity, whereas perceiving type prefers to keep their options open to alteration, likes impulsiveness and remains adaptable.

On the basis of these indicators, MBTI defines sixteen types to describe people's personalities, temperaments and approach towards general issues of life. For example, if an

────────────────
- *Arif Raza is with National University of Sciences and Technology, Pakistan. E-mail: arif_raza@mcs.edu.pk.*
- *Zaka Ul-Mustafa is with National University of Sciences and Technology Pakistan. E-mail: zaka@mcs.edu.pk.*
- *Luiz Fernando Capretz is with The University of Western Ontario, London, Canada. Email: lcapretz@uwo.ca*

individual is found to be the ESTJ type, it means that the individual prefers Extroversion, Sensing, Thinking, and Judging. This also signifies compatibility of personality types with a specific job and how one makes decisions in different situations. Although, these categories may uphold improved performance in certain situations, no category can be considered superior to other.

## 3 LITERATURE REVIEW

Choi, Deek and Im [3] study the effects of psychosocial factors of programmers' personality. In the study, a group of university students are first type profiled using the MBTI model. On the basis of MBTI type, they are split into alike, opposite and diverse (partially alike and partially opposite) groups. Each group is assessed for their output in code productivity. The subjects in the diverse MBTI type group show higher productivity than the other two. The work suggests that two novice programmers having similar MBTI dominant or auxiliary preferences (but not both), when paired together, would exhibit higher productivity level than those of other MBTI pair combination. However, many other factors which could influence the productivity such as code difficulty level, code quality measurement, subjects' programming experience and time limit to produce a code are not considered in the study.

Heinstrom [4] in her study considers five personality factors namely, neuroticism (to measure affect and emotional control), extraversion-introversion (a dimension that contrasts an outgoing character with a withdrawn nature), openness to experience (to measure depth, breadth and variability in one's imagination of experience), agreeableness (to compare caring and emotional support with competitiveness and hostility) and conscientiousness (to measure goal directed behavior and control over impulses). The research questions posed are about how each of these factors influences information behavior. The results show that information seeking could be linked to personality traits.

Bradley and Hebert [5] study the effect of personality type on team performance using a case example of two information system (IS) development teams that show a significant difference in their productivity. Although only two teams are studied, comparison demonstrates a good illustration of loss of productivity due to a poor combination of personality types. The study emphasizes that the team composition of personality types is an important parameter for differences in team performance. It further suggests that diversity and balance in team member personality types can lead to a successful team performance.

Sample [6] in his paper reviews the utility of MBTI from the perspective of organization development (OD) practitioner. Issues such as effects of personalities on communication and the conflict resolution, problem solving and decision making and team performances have been addressed.

Karn and Cowling [7] study the effects of different personality types using MBTI on the working of some software engineering (SE) team. The study describes how ethnographic methods could be used to study SE teams, to understand the role of human factors in a SE project. The results of the study indicate that certain personality types are more inclined to certain roles.

Capretz [8] reported that software engineers are a unique group of individuals. Although software engineering attracts people of all psychological types, certain traits are clearly more represented than others in this field. These findings do not mean that career success relates to the number of subjects of a type. As a matter of fact the software field is dominated by introverts, who typically have difficulty in communicating with the user. This may partially explain why software systems are notorious for not meeting users' requirements. Inspired by the MBTI, Capretz [9] has developed a range of practices for effective teaching and learning in a software engineering course. His aim is to reach every student, but in different ways, by devising various teaching approaches. As software engineering teachers tend to be ISTJ and INTJ and software engineering students ISTJ, this means that the teachers are reaching out the majority of their students. But the teaching of software engineering courses would be more effective for other types, such as Es and NFs if they incorporate and emphasize more open discussions and human factors issues. Feeling types like to see the personal implications of a concept. In a software engineering course this can be achieved with discussion on ethics and the human side of software management and team interaction.

Cecil [10] studied the personality types of professors teaching in information technology programs in the United States using the MBTI; and claimed that students are more likely to stay in computer science and IT program when there is a personality alignment between students and professors. Thus understanding personality type may aid computers science and software engineering programs to market, attract, and retain students.

All types choose software engineering, some types are more likely to stay within the field while others leave. Even so, software engineering is losing some atypical students who tried our wares and then sought more fitting studies; it means that we are losing some students of the types which can be important in transforming software engineering into a more user-oriented field and in finding new directions for software engineering in the future. If we can find ways to value the diversity among students, help them to go through the barrier of type and reach niches in software engineering where they will fit and feel valued, we should thrive to provide alternatives to retain them and enrich the profession [11].

Cunha and Greathead [12] in their study examine whether there is a specific personality type which could perform better at code reviewing. The code review task consists of 282 lines of Java code pattern search program which would operate on an ASCII file. 16 semantic bugs are inserted in the code of varying difficulty level. 64 undergraduate students are selected as subjects to complete the study. The results of the study reveal that NTs performed better than the other types on this task, on average.

Miller and Yin [13] in their research paper present a cognitive based mechanism for constructing software inspection teams and evaluate the process through an experiment to find out how it affects the inspection process. The subjects in the experiment are 33 graduate students, who are presented with an unseen document to inspect. After a series of experiments and analysis of results, the authors find themselves unable to demonstrate that the use of cognitive style information could benefit software inspection teams from effectiveness point of view.

Rutherfoord [14] in her experiment made use of Keirsey Temperament Sorter to select teams for a software engineering class, in which students can take the inventory and get their personality type results. The author observes that the groups having ESTJ personality types are very opinionated and follow a traditional path, unlike ISTJ group members, who are very quiet and private.

Karn and Cowling [15] also use MBTI to record the effect of personality type on behavior towards team members. They observe positive, negative and even both type of effects on a SE team. In another similar sort of research, the same authors observe two student teams to find out how individual personality types do the interaction during software development [16]. Other than the positive impacts, they also focus on possible disruptions that could occur.

Gifford et al. [17] in their research, study the relation of Management Team Roles – indicator (MTR-i), Belbin roles and MBTI for software teams. They find that these theories are related to each other and also study their impacts on software teams, too.

Greathead et al. [18] study the relationship between personality type and code review ability. Sixty four undergraduate students complete the study of reviewing 282 lines of Java code. However, only sensing scale and code review score are found to have a significant correlation.

Turley and Bieman [19] in their study identify the attributes differentiating exceptional and non-exceptional software engineers. After identifying the categories to be used in the quantitative analysis, they conduct an in-depth review of ten exceptional and ten non-exceptional software engineers working in a big computing organization, and use MBTI test results for the categorization of the data. Although no simple predictor of the performance is identified, they find MBTI results consistent with other studies, and observe most of the software developers exhibiting Introvert and Thinking personality types. They also identify 38 essential competencies of software engineers in the context of their importance towards their job.

Many empirical studies have also been carried out to explore the personality profiles of software professionals using MBTI tool. Bush and Schkade [20] survey 58 software professionals and conclude that ISTJ (25%) is the most common personality type, with INTJ (16%), and ENTP (9%) to follow. On the basis of the data collected from 47 scientific computer professionals Buie [21] also find out ISTJ (19%) as the most occurring personality type, followed by INTP (15%) and INTJ (13%). Smith [22] studies a sample of 37 systems analysts and reaches to the same conclusion of ISTJ (35%) as the most frequent type with ESTJ (30%) to follow. ISTJ (23%) is also found to be the most common type in Lyons [23] survey of 1229 software professionals from over 100 companies, INTJ (15%) to be the second, closely followed by INTP (12%).

## 4 RESEARCH METHODOLOGY AND ANALYSIS

In this study, we surveyed 18 SE professors and 92 SE students and of the National University of Sciences and Technology, Islamabad, Pakistan. A short version of the MBTI form (form G) was provided to identify their personality types. They were invited to take the MBTI measure at the university campus. The criteria to select the students to take part in this survey included their interest in software development projects as well as in taking MBTI test. Grade Point Averages (GPAs) of the students however were not taken into account. There were 18 professors (15 males, 3 females), 64 final-year under graduate (51 males, 13 females) students, and 28 post graduate (18 males and 10 females) students. The students' age range was between 21 and 23, whereas professors' age range was between 28 and 45 years old.

## 5 RESULTS AND ANALYSIS

The personality type distribution of faculty members is summarized in Table 1 below. It can be observed that among our respondents, introverts (72%) are more than double the extroverts (28%). Intuitive (56%) dominate over Sensing (44%), thinking (67%) over feeling (33%) and judging (56%) over perceiving (44%).

Table 1

Personality Type Distribution of Faculty Members in Each Dimension

| Personality Type | Percentage | Personality Type | Percentage |
|---|---|---|---|
| I | 72 % | E | 28 % |
| N | 56 % | S | 44 % |
| T | 67 % | F | 33 % |
| J | 56 % | P | 44 % |

We observe a slightly different outcome when these results are compared with some of the previous studies [1],[8],[24]. Our survey however confirms the over-representation of 'Ts' and 'Ns' and the under-representation of 'Fs' and 'Ss' respectively, as shown in Figure-1.

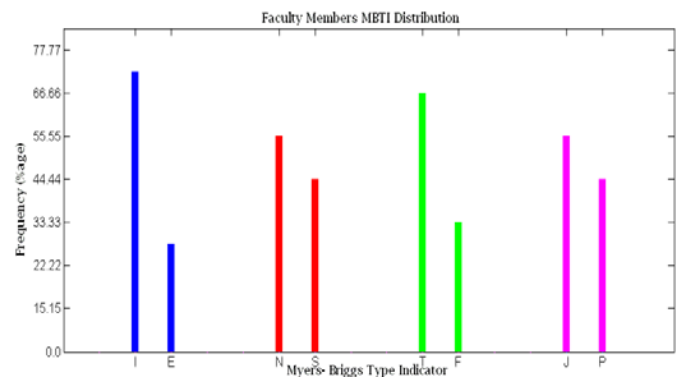

Fig.1. Faculty Personality Type Distribution

The personality type distribution of the students who took part in the survey is summarized in Table 2 below. It can be observed that among our respondents, introverts (55%) are more than extroverts (46%). Sensing (62%) dominate over intuitive (38%), thinking (59%) over feeling (41%) and perceiving (53%) over judging (47%).
The survey corroborates the over-representation of 'Ss' and 'Ts' and the under-representation of 'Ns' and 'Fs' respectively, as shown in Figure-2.

Unfortunately there is no registered MBTI personality type data available about the general Pakistani population to compare with our results, still this research indicates the pattern observed and existed among Pakistani software engineering professors and students.

Table 2

Students Personality Types Distribution in Each Dimension

| Personality Type | Percentage | Personality Type | Percentage |
|---|---|---|---|
| I | 55 % | E | 45 % |
| N | 38 % | S | 62 % |
| T | 59 % | F | 41 % |
| J | 46 % | P | 53 % |

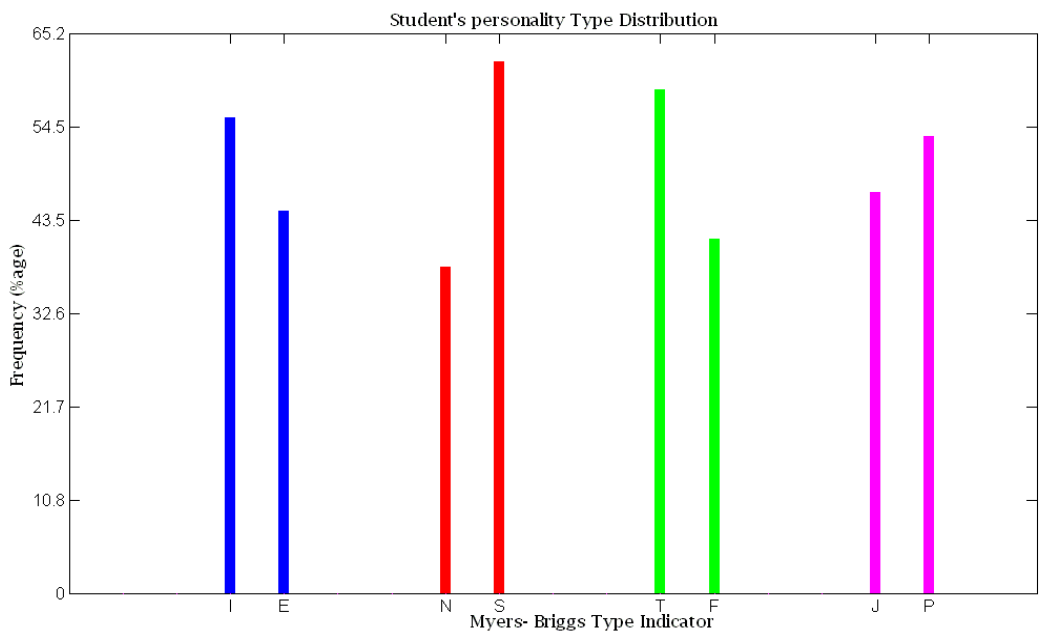

Fig. 2. Students' Personality Type Distribution

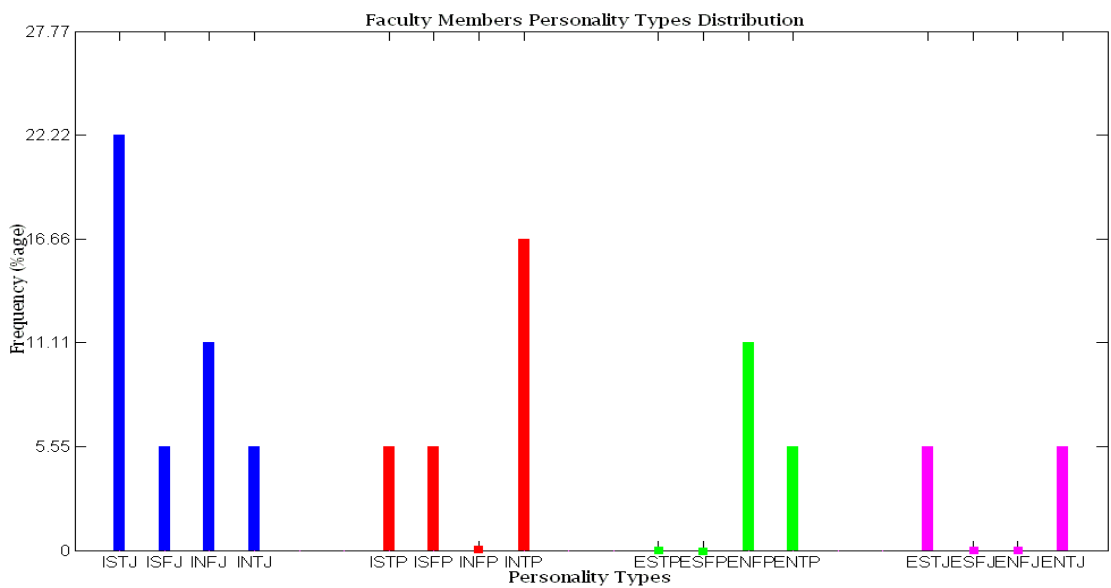

Fig.3. Faculty Members' Personality Type Distribution

Out of sixteen MBTI combinations, the ISTJ personality type has the top most representation of 22% among the surveyed Pakistani software engineering professors, as shown in Table 3.

Table 3

Faculty Members personality types distribution

| ISTJ 22% | ISFJ 5% | INFJ 11% | INTJ 5% |
|---|---|---|---|
| ISTP 5% | ISFP 5% | INFP 0 % | INTP 16.7% |
| ESTP 0 % | ESFP 0 % | ENFP 11% | ENTP 5% |
| ESTJ 5% | ESFJ 0 % | ENFJ 0 % | ENTJ 5% |

This is followed by INTP with a 17%, and then INFJ and ENFP both with 11%. Among the respondents, INFP, ESTP, ESFP, ESFJ and ENFJ have no representation, as presented in Figure -3

ISTJ personality type has got the top most representation among the surveyed Pakistani software engineering students too, along with ENTP with 12%, as shown in Table 4.

Table 4

SE Students personality types distribution

| ISTJ 12 % | ISFJ 10% | INFJ 1% | INTJ 8% |
|---|---|---|---|
| ISTP 7% | ISFP 8% | INFP 4% | INTP 7% |
| ESTP 4% | ESFP 8% | ENFP 4% | ENTP 12% |
| ESTJ 10% | ESFJ 4% | ENFJ 2% | ENTJ 0 % |

This is followed by ISFJ and ESTJ with a 10%, and then INTJ, ISFP, and ESFP with 8%. Among the respondents, ENTJ has no representation, while INFJ (1%) and ENFJ(2%) are also under represented, as represented in Figure 4.

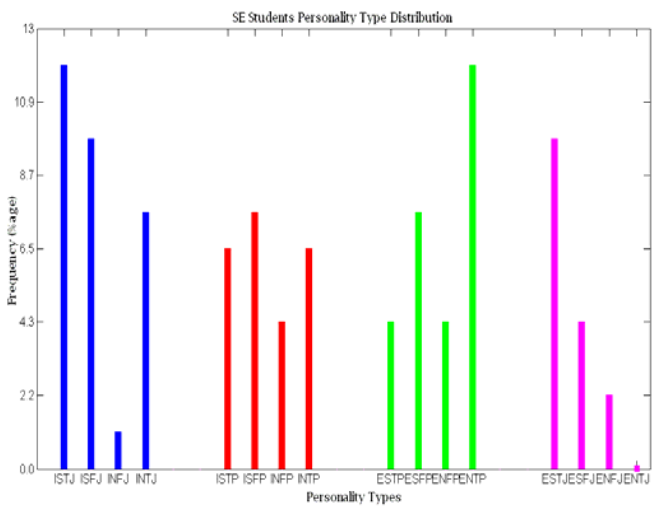

Fig.4. Students personality Type distribution

The sampled Pakistani software engineering faculty members' temperament distribution has also been recorded and is reflected in Table 5. The dominant temperament is IT with 50%, although IJ (44%), IS (39%), SJ (33%), NT (33%), IN(33%), ST (33%) and NP (33%) are well represented as well. ES is the least represented temperament with only 6%, as shown in Table 5. These values, however, are not very similar to the results of previous studies [4], [6], [9], [14], [20], [23] where STs and TJs have been marked as abundant and NFs as scarce.

Table 5

Faculty Members Temperament Distribution

| Temperament | Percentage (%) | Temperament | Percentage (%) |
|---|---|---|---|
| SP | 11 | TJ | 29 |
| SJ | 33 | TP | 28 |
| NT | 33 | FP | 17 |
| NF | 17 | FJ | 17 |
| IJ | 44 | IN | 33 |
| IP | 28 | EN | 22 |
| EP | 17 | IS | 39 |
| EJ | 11 | ES | 6 |
| ST | 33 | ET | 17 |
| SF | 11 | EF | 11 |
| NP | 33 | IF | 22 |
| NJ | 22 | IT | 50 |

This study indicates ITs (Introverts and Thinking) and IJs (Introverts and Judging) are the leading temperaments among studied professors. According to the MBTI stipulations, ITs prefer to work alone and their decisions are based on a logical and objective analysis, whereas IJs manifest in an orderly manner.

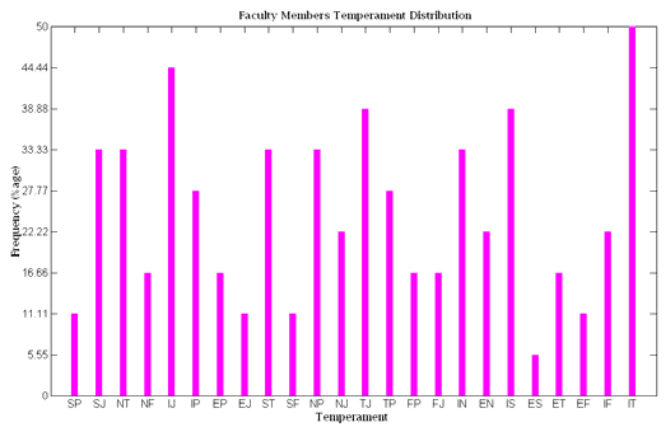

Fig.5. Faculty Temperaments Distribution

According to the observed results of the sampled Pakistani software engineering students' temperament distribution, as reflected in Table 6, SJ and IS are found to be the dominant temperament with 36% representation. Other well represented temperaments include ST (33%), IJ (30%), SF (29%), TJ (29%), and TP (29%). NF is the least represented temperament with only 12%, as shown in Table 6, in line with the results of previous studies [8], [13], [22].

Table 6
Students Temperament Distribution

| Temperament | Percentage (%) | Temperament | Percentage (%) |
|---|---|---|---|
| SP | 26 | TJ | 29 |
| SJ | 36 | TP | 29 |
| NT | 26 | FP | 24 |
| NF | 12 | FJ | 17 |
| IJ | 30 | IN | 20 |
| IP | 25 | EN | 18 |
| EP | 28 | IS | 36 |
| EJ | 16 | ES | 26 |
| ST | 33 | ET | 26 |
| SF | 29 | EF | 18 |
| NP | 27 | IF | 23 |
| NJ | 11 | IT | 33 |

Among the students' data, the results of the study show SJs (Sensing and Judging) and ISs (Introverts and Sensing) as the dominant temperaments. SJs are attuned to the practical, hands-on, common-sense view of events and tend to seek closure, and want things settled. ISs tend to draw energy from the internal world of ideas, and attuned to be organized. ISTJ and ENTP are the prevailing personality types.

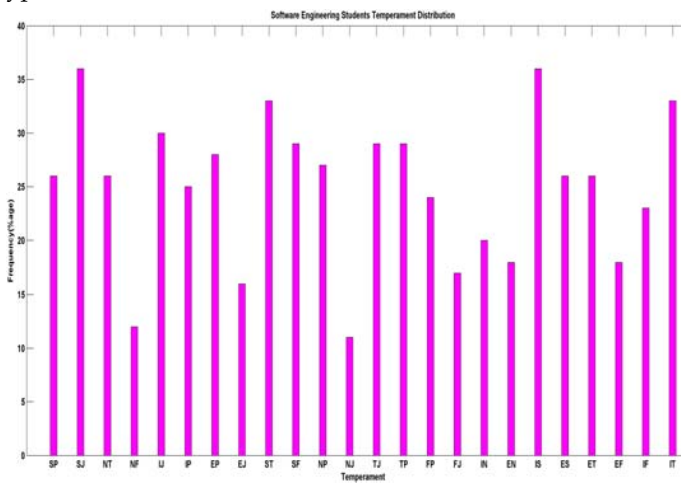

## 6 DISCUSSION

In our study, although software engineering faculty members and students share many similarities in their personality type distributions, there are also some differences between these two groups. For instance, ISTJ configurations predominate in both the samples. Conversely INFP, ESTP, ESFP, ESFJ and ENFJ have no representation among professors, whereas the least represented types among students are ENTJ, INFJ and ENFJ; they all are of intuitive (N) and judging (F) nature. The most prominent discrepancies between the two groups occur in the following types: ENTP, accounts for 5% of the professors as opposed to 12% of the students. Similarly both ISFJ and ESTJ are demonstrated in 5% of faculty members versus 10% of the students, and ESFP, which is shown in 8% of the students, has no representation among professors.

It is also vital to examine the behavior in the different dimensions. In the introvert/extrovert (IE) dimension, both the professors and the students are more introverts than extrovert. Introvert/Extrovert proportion, thus, in both the samples support the historical studies that the software developers are mainly introverts. Similarly, thinking people outnumber feeling individuals in both the samples, with 67% and 59% of professors and students respectively. However, there are significant differences between the two groups within the intuitive /sensing (NS) scale, where more professors are intuitive (56%) than sensing (44%). On the other hand, students tend to be more sensing (62%) as compared to intuitive (38%). Similarly, among professors, judging (56%) dominate over perceiving (44%); whereas in the students' sample, perceiving (53%) outnumber judging (47%). The implications of these discrepancies highlight personality differences among teachers and students.

Among the sample engineering faculty members IT is the dominant temperament with IJ, NT, IN, and NP are well represented; whereas among surveyed students SF, TJ, and TP are included in well represented temperaments. This is to be noted that IJ, IS, SJ, and ST are well represented in both the samples. On the other hand, ES is the least represented temperament among professors and NF is the least represented temperament among students. There are almost three times as many students who prefer sensing and perceiving as there are faculty members with this combination. SP students prefer a flexible approach to factual material. Their NJ professors, on the other hand, prefer structure and theories. The SP student are more likely to view the facts themselves as more important than the theories and are less likely to want the facts organized according to some grand structure. Furthermore, no matter which preference combination you look at, it is clear that sensing types will probably need to learn to cope with the intuitive environment preferred by the majority of their professors.

## 7 CONCLUSIONS

Although all personality types contribute towards problem solving one way or the other, Capretz [9] states that software engineering discipline attracts people of all psychological types, even though certain personalities have more representation than others in this field. According to our survey analysis, both the professors and the students tend to be introverts. Similarly, in both the samples the dominant personality type is a combination of Introversion, Sensing, Thinking, and Judging (ISTJ). However, another observation which is evident in the comparative analysis between professors and students of software engineering indicate that ITs (Introverts and Thinking) and IJs

(Introverts and Judging) are the dominant temperaments among studied professors, whereas SJs (Sensing and Judging) and ISs (Introverts and Sensing) are the leading temperaments among students.

In closing, software engineering has, and will continue to have, the challenge of engaging the interest, at the same time, of those students whose minds work in a linear fashion (S) and of those whose mind concern themselves with patterns (N). Software engineering programs cannot afford losing types who can do software engineering well – the practical, hands-on linear thinkers (ST types). To retain these students, teaching needs to be very clear, and sequential, with explicit practical applications. This kind of teaching is often seen as too slow by the fast-moving intuitive students, especially those combining extroversion with intuition (EN) types. EN types are often found in top management in the information technology industry. They would be our most innovative and action-oriented colleagues. Teachers can motivate and challenge EN types by giving the future vision or the big picture and by assigning group projects involving integration of complex software systems.

## 8 REFERENCES

[1] Bishop-Clark C. and Wheeler, D.D. "The Myers Briggs Personality Type and Its Relationship to Computer Programming," *Journal of Research on Computing in Education*, Vol. 26, no. 3, pp. 358-370, 1994.

[2] Myers I.B., McCaulley M.H., Quenk N.L. and Hammer A.L., "MBTI Manual," *A Guide to the Development and Use of the Myers-Briggs Type Indicator,* Palo Alto, California, Consulting Psychologists Press, 1998.

[3] Choi K. S., Deek F. P., Im I., "Exploring the Underlying Aspects of Pair Programming: The Impact of Personality", *Journal of Information and Software Technology*, Vol. 50 no. 11, pp. 1114-1126, 2008

[4] Heinstrom J., "Five Personality Dimensions and Their Influence on Information Behavior", *Information Research*, Vol. 9 no.1, 2003

[5] Bradley J. H. and Hebert F. J., "The Effect of Personality Type on Team Performance", *Journal of Management Development,* Vol. 16 no. 5, pp. 337-353, 1997

[6] Sample J., "The Myers-Briggs Type Indicator and OD: Implication for Practice from Research", *Organization Development Journal*, Vol. 22, no 1, pp. 67-75, 2004

[7] Karn J.S and Cowling A.J., "Using Ethnographic Methods to Carry Out Human Factors Research in Software Engineering", *Behavior Research Methods*, Vol. 38 no. 3, pp. 495-503, 2006

[8] Capretz L.F., "Personality Types in Software Engineering", *Int. J. Human-computer Studies*, Vol. 58, no. 2, pp. 207-214, 2003

[9] Capretz L.F., "Implications of MBTI in Software Engineering Education", *SIGCSE Bulletin*, Vol.4 no.4, pp. 134- 137, 2002

[10] Cecil, D.K. "Personality Types of IT Professors". *ACM SIGITE, Fairfax, Virginia*, pp.13-23, 2009

[11] Capretz L.F. and Ahmed F. "Making Sense of Software Development and Personality Types", *IEEE IT Professional*, Vol. 12, no. 1, pp. 6-13, IEEE Press, Doi: 10.1109/MITP.2010.33

[12] Cunha A. Devito Da and Greathead D., "Code Review and Personality: Is Performance Linked to MBTI Type?." University of Newcastle Upon Tyne, 2004.

[13] Miller J. and Yin Z., "A Cognitive-Based Mechanism for Constructing Software Inspection Teams", *IEEE Transactions on Software Engineering*, Vol.30, no.11, pp. 811 – 825, 2004

[14] Rutherfoord R. H., "Using Personality Inventories to Help Form Teams for Software Engineering Class Projects", Annual Joint Conference Integrating Technology into Computer Science Education, *Proceedings of the 6th annual conference on Innovation and technology in computer science education, Canterbury, United Kingdom*, pp. 73-76, 2001

[15] Karn, J.S. and Cowling, A.J. "An Initial Study of the Effect of Personality on Group Projects in Software Engineering", Research Report, Department of Computer Science CS-04-01, University of Sheffield, pp. 1-49, 2001

[16] Karn, J.S. and Cowling, A.J. "An Initial Observational Study of the Effects of Personality Type on Software Engineering Teams. *Proceedings of the 8th International Conference on Empirical Assessment in Software Engineering (EASE), IEE, Edinburgh, UK*, 2004

[17] Gifford, S., Henry, S., Klaus P., Hollander, S. "Team Assembly for Virtual Corporations", *First Seminar on Advanced Research in* Electronic Business, pp. 94-104, 2001

[18] Greathead, D., Arief, B., Coleman, J. "TA Weinberg", *Proceedings of the 5th annual DIRC Research Conference, Mackie, J., Rouncefield, M.*, pp. 25-29, 2005

[19] Turley, R. T. and J. M. Bieman "Competencies of Exceptional and Non-Exceptional Software Engineers," *Journal of Systems and Software* Vol.28 no.1, pp. 19-39, 1995

[20] Bush C.M. and Schkade LL. "In Search of the Perfect Programmer", *Datamation*, Vol. 31 no. 6, pp. 128 132, 1985

[21] Buie E.A. " Psychological Type and Job Satisfaction in Scientific Computer Professionals". *Journal of Psychological Type*, Vol. 15, pp. 50-53, 1988

[22] Smith D.C.. "The Personality of The Systems Analyst: An Investigation. *SIGCPR Computers*, Vol. 12 no.2, pp. 12-14, 1989

[23] Lyons M. "The DP Psyche". *Datamation*, Vol. 31 no. 16, pp. 103-110. 1985

[24] Capretz LF. "Psychological Types of Brazilian Software Engineering Students", *Journal of Psychological Type,* Vol. 68 no. pp. 5-37, 2008.

[25]Hignite, M.A., Saltzinger, J.W. and Margavio, T., Correlated factors of success in information systems: personality, creativity, and academic achievement." *Proceedings of the Americas Conference on Information Systems AMCIS.*, Baltimore, United States, 1998

[26] Seddigi Z.S, Capretz LF, House D. "A Multicultural Comparison of Engineering Students: Implications to Teaching and Learning", *Journal of Social Sciences,* Vol. 5 no. 2, pp.117, 2009

**AUTHORS' BIOGRAPHIES**

Arif Raza received his M.Sc. (1994) in Computing Science from University of London (U.K.), and PhD (2011) in Software Engineering from the University of Western Ontario (Canada). During his professional career, he has been actively involved in teaching and research. Dr. Raza has authored and co-authored several research articles in peer reviewed journals and conference proceedings. His current research interests include usability issues in open source software (OSS), human computer interaction, human factors and empirical studies in software engineering. He is presently serving as Assistant Professor at University of Sciences and Technology, Pakistan and can be reached at arif_raza@mcs.edu.pk.

Zaka Ul-Mustafa has been in the field of academia for last nine years. He has been providing consultancy to many industrial organizations and enterprises. He did his under-graduation in Electrical Engineering and post graduation in Software Engineering from National University of Sciences and Technology Pakistan. He has many research publications to his credit. His current research areas include Open Source Software Usability, Unified Modeling Language, and Wireless Networks. He can be contacted at zaka@mcs.edu.pk

Luiz Fernando Capretz has over 30 years of international experience in the software engineering field as a practitioner, manager and educator. Having worked in Brazil, Argentina, U.K., Japan, Italy and the UAE, he is currently Assistant Dean (IT and e-Learning) at the University of Western Ontario, Canada. He has published over 100 peer-reviewed research papers on software engineering in leading international journals and conference proceedings, and he has co-authored two books in the field. His present research interests include software engineering (SE), human factors in SE, software estimation, software product lines, and software engineering education. Dr. Capretz received his Ph.D. in Computing Science from the University of Newcastle upon Tyne (U.K.), his M.Sc. in Applied Computing from the National Institute for Space Research (INPE, Brazil), and his B.Sc. in Computer Science from State University of Campinas (UNICAMP, Brazil). He is an IEEE senior member, ACM member, MBTI certified practitioner, Professional Engineer in Ontario (Canada), and he can be contacted at lcapretz@uwo.ca.